# Kondo Effect in Mesoscopic Quantum Dots

*M. Grobis,[1] I. G. Rau,[2] R. M. Potok,[1,3] and D. Goldhaber-Gordon[1]*

1       Department of Physics, Stanford University, Stanford, CA 94305
2       Department of Applied Physics, Stanford University, Stanford, CA 94305
3       Department of Physics, Harvard University, Cambridge, MA 02138

## Abstract

A dilute concentration of magnetic impurities can dramatically affect the transport properties of an otherwise pure metal. This phenomenon, known as the Kondo effect, originates from the interactions of individual magnetic impurities with the conduction electrons. Nearly a decade ago, the Kondo effect was observed in a new system, in which the magnetic moment stems from a single unpaired spin in a lithographically defined quantum dot, or artificial atom. The discovery of the Kondo effect in artificial atoms spurred a revival in the study of Kondo physics, due in part to the unprecedented control of relevant parameters in these systems. In this review we discuss the physics, origins, and phenomenology of the Kondo effect in the context of recent quantum dot experiments. After a brief historical introduction (Sec. 1), we first discuss the spin-½ Kondo effect (Sec. 2) and how it is modified by various parameters, external couplings, and non-ideal conduction reservoirs (Sec. 3). Next, we discuss measurements of more exotic Kondo systems (Sec. 4) and conclude with some possible future directions (Sec. 5).

## Keywords

# 1     Introduction

At low temperatures, a small concentration of magnetic impurities — atoms or ions with a non-zero magnetic moment — can dramatically affect the behavior of conduction electrons in an otherwise pure metal. This phenomenon, known as the Kondo effect (Kondo, 1964), has been a leitmotif of solid-state physics since the 1960s. Nearly a decade ago, the Kondo effect was discovered in a new system (Goldhaber-Gordon et al., 1998a), in which the local magnetic moment belongs not to an atom but to a lithographically-defined droplet of electrons known as a quantum dot or artificial atom (Kastner, 1993). When the droplet contains an odd number of electrons, it has a net spin and hence may be thought of as a magnetic artificial atom. Nearby metal or semiconducting electrical leads play the role of the host metal.

The artificial atom system offers several important advantages for the study of Kondo effect:

1. The microscopic state of the system is better-defined than in bulk metals and may even be tailored by varying the system's geometry.
2. Almost all the important parameters may be precisely measured and tuned.
3. A single impurity site is measured, rather than a statistical average over many sites.
4. The local site may be studied out of equilibrium.

In this chapter of the Handbook we review how researchers around the world have applied these advantages to create a renaissance in the study of the Kondo effect. Though this approach was initially inspired by theory (Glazman and Raikh, 1988, Ng and Lee, 1988) and continues to be strongly informed by theory, we organize our review around experimental developments in mesoscopic systems.

## 1.1     Background of Kondo Effect: Dilute Magnetic Alloys

At relatively high temperatures, the resistivity of a metal is dominated by electron-phonon scattering. As the temperature is lowered well below the Debye temperature the scattering rate, and hence the phonon contribution to resistivity, decreases as $T^5$. At sufficiently low temperatures phonon-induced scattering becomes insignificant and resistivity saturates at a finite value determined by scattering from defects in the crystal lattice. In the 1930s, researchers noticed that this simple picture did not always hold. Measurements of Au cooled below 10 K sometimes showed a resistivity rise rather than saturation as the temperature was lowered further (de Haas et al., 1934). The effect remained an enigma until the 1960s when experimental evidence correlated the low-temperature resistivity rise with the presence of dilute magnetic impurities in the metal (Sarachik et al., 1964). Inspired by the strong evidence that individual magnetic impurities were responsible for the resistivity rise, J. Kondo considered a model involving an antiferromagnetic interaction between the local moment and the sea of conduction electrons (Kondo, 1964). Using perturbation theory, Kondo found that the antiferromagnetic interaction leads to a logarithmic rise in electron-impurity scattering with decreasing temperature.

Though Kondo's work was a crucial breakthrough, calculations using this theoretical framework inherently produced unphysical logarithmic divergences. Finding a solution to this conundrum became known as the "Kondo Problem." The problem was finally solved in 1975 by K. Wilson, who developed a new renormalization group (RG) technique for the purpose (Wilson, 1975). Wilson's RG calculations showed that at temperatures below a characteristic Kondo temperature $T_K$ a magnetic impurity would form a singlet with the surrounding sea of conduction electrons. Despite this success in determining the ground state (and thermodynamic properties) of a magnetic impurity in a metal, nearly two decades would pass before the development of numerical RG techniques that could accurately calculate transport properties such as resistivity over a broad range of temperatures (Costi et al., 1994).

## 1.2   Tunneling: Kondo in a New Geometry

Since the Kondo effect is a property of magnetic impurities in a host metal, it should also be observable in transport through magnetic impurities sandwiched between two metal leads (Appelbaum, 1966). This is borne out in both large tunnel junctions with many impurities (Wyatt, 1964) and more recently in a nanometer-scale junction containing a single impurity (Gregory, 1992, Ralph and Buhrman, 1994). In this geometry, the Kondo effect enhances conductance rather than resistivity at low temperature and low bias voltage (Appelbaum, 1966).

The advent of high mobility GaAs heterostructures in the 1980s ushered in the modern era of mesoscopic semiconductor physics: the study of electronic phenomena at intermediate spatial scales, between atomic and macroscopic. Through concurrent advancement in lithographic technology, the first artificial atoms were created in patterned GaAs heterostructures in the late 1980s (Reed et al., 1988, van Wees et al., 1989). Around the same time, two theoretical teams pointed out that this type of system can behave like a magnetic impurity and hence should exhibit the Kondo effect (Glazman and Raikh, 1988, Ng and Lee, 1988). The diverse experiments on the Kondo effect in such semiconductor nanostructures are the main focus of this review. However, these are not the only new and interesting Kondo systems. Electrons localized on a single molecule or a single carbon nanotube can exhibit Kondo effects stemming from local degeneracies not easily achieved in a GaAs artificial atom (Park et al., 2002, Liang et al., 2002), and can be coupled to reservoirs with exotic properties (superconducting, ferromagnetic). The Kondo effect has also been observed in STM measurements of surface adatoms (Madhavan et al., 1998, Li et al., 1998) and metal complexes (Zhao et al., 2005), allowing (for example) the study of the spatial characteristics of Kondo interactions (Manoharan et al., 2000).  Though a wide range of Kondo systems have played important roles in developing our understanding of the Kondo effect, we will draw mainly on work focusing on mesoscopic systems.

## 2 Experimental Signatures of Spin-½ Kondo Effect

In this section we review the characteristic signatures of the Kondo effect in the context of a spin-½ quantum dot coupled to two reservoirs. Consideration of more complex scenarios is left to later sections.

### 2.1 Quantum Dots

Our discussion of experimental results will draw primarily on work performed on three types of quantum dots, each offering a different set of advantages: lateral and vertical semiconductor dots, and carbon nanotube quantum dots (Fig. 1). Lateral quantum dots are formed by depleting a sub-surface two dimensional electron gas (2DEG) using lithographically-defined metallic gates, as shown in Fig. 1(a). The 2DEG resides at the interface of semiconductor heterostructure (e.g. GaAs/AlGaAs), located tens to hundreds of nanometers below the surface. The quantum dot consists of a pool of confined electrons, tunnel coupled to extended sections of the 2DEG, which serve as leads. Several nearby gate electrodes electrostatically control the quantum dot electron occupancy, as well as the dot-lead couplings. In principle the number of electrons can be reduced to zero, however in many lateral quantum dot geometries the conductance through the quantum dot's contacts is "pinched off" by the increasingly-negative gate voltage before the dot is entirely emptied.

The vertical quantum dot geometry, shown in Fig. 1(b), allows the formation of a small, well-coupled, few-electron quantum dot whose conductance remains measurable all the way down to the expulsion of the last electron. This allows accurate determination of the electron occupancy, as discussed later in this section. Another important difference between lateral and vertical quantum dots is the number of tunneling modes, or channels, that couple the dot to the leads.[1] Typically, the dot-lead contacts in lateral dots are single-mode quantum point contacts (QPCs), while the wide lead-dot contact area in vertical dots contains several partially-transmitting modes.

A carbon nanotube quantum dot is created using a single carbon nanotube deposited on a conducting substrate covered by a thin insulator (e.g. $SiO_2$ / $n^+$-Si). Lithographically-defined metal leads contact the nanotube, and the conducting substrate is biased to modify the electron occupancy (Fig. 1(c)). While the nanotube-lead couplings cannot be finely controlled, this geometry allows for the creation of wide variety of leads, such as superconducting and magnetic, which are difficult, if not impossible, to achieve in semiconductor heterostructures. For a further discussion of quantum dots, including the Kondo effect, we encourage readers to examine several recent reviews (Kouwenhoven et al., 1997, Kouwenhoven et al., 2001, Yoffe, 2001, Pustilnik and Glazman, 2004, Giuliano et al., 2004).

### 2.2 Theoretical Background

We will present only a brief theoretical description of the Kondo effect. The reader may consult several references for a more complete theoretical description (Gruner and Zawadowski, 1974, Wilson, 1975, Nozières and Blandin, 1980, Andrei et al., 1983, Hewson, 1993, Costi et al., 1994, Pustilnik and Glazman, 2004).

---

[1] Though the terms "tunneling channels" and "tunneling modes" are often used interchangeably, we will avoid using former to avoid confusion with the concept of independent Kondo screening channels, which lead to the multi-channel Kondo effect (Sec. 5.2).

### 2.2.1 Kondo Hamiltonian

The Kondo effect arises when a degenerate local state is coupled to a reservoir of mobile electrons. The system is well described by the Anderson Hamiltonian

$$H = \sum_{k,\sigma} \varepsilon_k c_{k\sigma}^\dagger c_{k\sigma} + \sum_\sigma \varepsilon_0 d_\sigma^\dagger d_\sigma + U d_\uparrow^\dagger d_\uparrow d_\downarrow^\dagger d_\downarrow + \sum_{k,\sigma} \left( v_k d_\sigma^\dagger c_{k\sigma} + v_k^* c_{k\sigma}^\dagger d_\sigma \right), \tag{1}$$

which was originally proposed to describe a magnetic impurity atom in a metal (Anderson, 1961). The first term represents the kinetic energy of electrons in the reservoir, labeled by their momentum and spin. The second term is the quantized energy of localized electrons in a single spin-degenerate state near $E_F$ – all other quantum dot levels are assumed to be either completely full (well below $E_F$) or completely empty (well above $E_F$), and hence can be safely ignored. The third term accounts for interactions among localized electrons: a second electron added to the local site costs more than the first electron, by an amount known as the charging energy $U$. This is crucial, since it allows for the possibility that the doubly-degenerate site can be occupied by just a single electron. In this case, the site acts as a magnetic impurity, and can exhibit Kondo effect. The final term describes spin-conserving tunneling on and off the local site, with strength $v_k$. A schematic of the Anderson model is shown in Fig. 2. A spin-degenerate quantum dot is probably the most exact experimental realization of the Anderson model.

With a few simple assumptions, which are generally well-justified for a quantum dot with odd occupancy, the interaction between the conduction electrons and the quantum dot (final term in equation (1)) can be approximated as an antiferromagnetic coupling,

$$H_{\text{int}} = J s_{cond} \cdot S_{QD}. \tag{2}$$

Equation (2) is known as the s-d or Kondo Hamiltonian (Kondo, 1964, Schrieffer and Wolff, 1966, Kasuya, 1956, Yosida, 1957) as alluded to previously. $S_{QD}$ is the net spin of the quantum dot, while $s_{cond}$ is sum of spin operators for the conduction electrons (see (Pustilnik and Glazman, 2004) for more details). The strength of the antiferromagnetic coupling depends on the microscopic parameters of the Anderson model, according to

$$J \sim |v|^2 \left( \frac{1}{E_F - \varepsilon_0} + \frac{1}{\varepsilon_0 + U - E_F} \right). \tag{3}$$

Here, the interaction strength is taken to be a constant ($v_k = v$) over a finite bandwidth ($D$) and zero otherwise, as indicated in Fig. 2.

### 2.2.2 Kondo Ground State

The ground state of the Kondo Hamiltonian is not two-fold degenerate, but is instead a spin singlet in which the spin of a localized electron is matched with the spin of delocalized electrons to yield a net spin of zero. This "Kondo screening cloud" is shown schematically in Fig. 3. Due to phase space constraints, the Kondo interaction predominantly affects electrons at the Fermi surface. As a result, the spectroscopic signature of the Kondo singlet is the formation of narrow resonance energetically pinned to the Fermi energy.[2]

Due to its non-conventional nature, the "binding energy" of the Kondo singlet is not simply proportional to $J$. Since the formation of the Kondo singlet is a continuous transition, the binding energy is typically expressed in the form of a Kondo temperature $T_K$. This cross-over energy scale can be calculated perturbatively using equation (2) to give,

$$T_K \sim D\sqrt{\rho J}\exp\left(-1/2\rho J\right),\tag{4}$$

where $\rho$ is the density of states in the reservoir at energy $E_F$ (Hewson, 1993). It is important to realize that $T_K$ is not a well-defined energy scale: equally-valid definitions can differ by a constant multiplicative factor. In this review we will use a definition suggested by Theo Costi (Costi et al., 1994) and introduced for quantum dots by Goldhaber-Gordon et al. (1998b): $T_K$ is the temperature at which the Kondo conductance has risen to half its extrapolated zero-temperature value (see Sec. 2.3.3).

The expression for $T_K$ can be rewritten in terms of quantities that are more easily controlled and measured in quantum dot experiments (Haldane, 1978):

$$T_K \sim \sqrt{\Gamma U}\exp\left(\pi\varepsilon_0(\varepsilon_0+U)/\Gamma U\right).\tag{5}$$

Here, $\varepsilon_0$ is the energy of the localized state measured relative to $E_F$ and $\Gamma$ is the rate for electron tunneling on and off the dot. The spin-½ Kondo effect occurs only when the local site is singly occupied ($\varepsilon_0 < 0$ and $\varepsilon_0 + U > 0$), so the exponent in equation (5) is negative, as expected from equation (4). These equations break down near the charge degeneracy points ($\varepsilon_0 = 0$ and $\varepsilon_0 + U = 0$), where charge fluctuations produce mixed-valence rather than Kondo behavior (Costi et al., 1994). Coupling additional leads to the quantum dot modifies $\Gamma$, but does not affect overall Kondo behavior. The leads behave as a single collective reservoir, as long as Kondo processes can freely exchange electrons between each pair of reservoirs (Glazman and Raikh, 1988).

$T_K$ is maximized when the quantum dot state is energetically close to $E_F$ ($\varepsilon_0 \sim 0$) and strongly coupled to the leads (large $\Gamma$). Finite charging energy ($U$) and level spacing ($\Delta$) put a limit on $T_K$ in quantum dots, since electrons are no longer localized on the dot for $\Gamma > \min(U, \Delta)$. $U$ and $\Delta$ increase with decreasing device size, yielding a maximal $T_K$ in the range of 0.1-1 K for typical semiconductor quantum dots and up to 10 K in carbon nanotube based devices. For comparison, $T_K$ in excess of 500 K has been observed for atomic impurities (Gruner and Zawadowski, 1974).

The formal spatial scale of the Kondo cloud is given by

$$\xi_K \sim \hbar v_F / k_B T_K,\tag{6}$$

where $v_F$ is the Fermi velocity (Sorensen and Affleck, 1996). $\xi_K \sim 1\ \mu m$ for typical semiconductor quantum dots in the Kondo regime, which gives an estimate of the spatial extent of Kondo correlations. The spatial properties of the Kondo cloud have proved difficult to measure. This is understood to stem from oscillations and polynomial decay that modulate the slow exponential decay of spin polarization in the electron gas at distances greater than $\lambda_F$ from the local moment (Barzykin and Affleck, 1996). In addition, charge rearrangement due to Kondo singlet formation is minimal and occurs only at length scales on the order of $\lambda_F$. Local probes, such as STM, generally measure the total density of states, which is modified on similar short spatial scales (Ujsaghy et al., 2000).

---

[2] In the mixed-valence regime the Kondo resonance is expected to move away from $E_F$ and no longer be symmetric about its maximum (Costi et al., 1994, Wahl et al., 2004).

## 2.3    Kondo Transport in Quantum Dots

Formation of the Kondo screening cloud enhances the scattering of conduction electrons by the local magnetic site. Hence, the resistivity of a bulk metal with a small concentration of magnetic impurities rises as temperature is decreased below the characteristic Kondo temperature for that impurity/metal system. In a geometry where transport is dominated by tunneling through a magnetic site, the enhanced scattering has the opposite effect: *conductance* rises for $T < T_K$, as a new mechanism for transport becomes available (Appelbaum, 1966, Ng and Lee, 1988, Costi et al., 1994).  The Kondo effect in transport through semiconductor quantum dots was first reported in 1998 (Goldhaber-Gordon et al., 1998a, Cronenwett et al., 1998, Simmel et al., 1999), and has since been observed and investigated further by research groups worldwide. As the occupancy of a dot is tuned by a nearby gate, the low-temperature conductance alternates between high (odd occupancy) and low (even occupancy), as seen in Fig. 4(a). The high conductance for odd occupancy disappears as the coupling to the leads ($\Gamma$) is reduced (Fig. 4(b)): $T_K$ becomes lower than the base temperature of the measurement apparatus (*cf.* equation (2.5)).  In this "Coulomb blockade" (CB) regime, conduction only occurs at the charge degeneracy points, whose extrapolated zero-temperature width is proportional to $\Gamma$.

Kondo-mediated transport through a quantum dot may be distinguished from other mechanisms for conductance enhancement by the following distinctive features:

1) Even-odd alternation of conductance as a function of dot occupancy (Kondo effect occurs only for odd occupancy)
2) Enhanced conductance at low-temperature, with a characteristic temperature dependence
3) Enhanced conductance at zero bias relative to finite bias
4) Suppression of conductance by magnetic field, which splits the local spin degeneracy
5) Recovery of enhanced conductance at finite magnetic field by application of a bias voltage equal to the Zeeman splitting

The above list is useful for demonstrating the occurrence of a Kondo effect, but is not an exhaustive or quantitative description of spin-½ Kondo. Note that some of the listed features can only be observed in systems where the local site occupancy can be tuned (i.e (1)) or where a finite bias can be applied across the local site(s) (i.e. (3) and (5)).

### 2.3.1    Even-Odd Effect

The low temperature Kondo conductance enhancement is expected to occur whenever a local state is degenerate and partially filled.  In the absence of additional degeneracies, this occurs whenever the number of electrons on the dot is odd.  Though early Kondo dot measurements exhibited conductance enhancements in alternating

Coulomb blockade valleys, the researchers could not directly demonstrate that the enhancement occurred for odd rather than even occupancies, since the initial occupancy of the dot was unknown (Goldhaber-Gordon et al., 1998a). Indeed, in later sections we discuss a variety of Kondo effects that do not require a spin-½ dot state and hence do not follow the even-odd rule. More recent quantum dot devices use either of two methods to conclusively determine the total electron occupancy. In one method, the electron number is determined by counting Coulomb blockade peaks as the quantum dot is emptied by an increasingly negative gate voltage. This technique works best in vertical quantum dots (Tarucha et al., 1996) (Fig. 1(b)), as well as certain lateral dots (Ciorga et al., 2000), which remain conducting as the occupancy is reduced and counted down to zero. A second technique uses a quantum point contact (QPC) fabricated close to the quantum dot, to serve as a detector of the dot's occupancy (Sprinzak et al., 2002). Each electron added to the quantum dot electrostatically alters the conductance of the QPC. This method works even if transport through the quantum dot is so slow as to be immeasurable with conventional methods (fewer than around 1000 electrons per second.) These methods confirm that in small GaAs quantum dots spin-½ Kondo and even-odd conductance alternation are the rule rather than the exception.

### 2.3.2 Dependence on External Parameters

Kondo conductance through a quantum dot has a characteristic and calculable dependence on external parameters such as temperature, magnetic field, and voltage across the dot (Pustilnik and Glazman, 2004). We can consider two regimes: low energy and high energy, distinguished by whether the perturbations $k_B T$, $eV$, and $g\mu_B B$ are small or large compared to $k_B T_K$. At low energy, a Kondo impurity acts as an elastic scatterer, and nearby conduction electrons behave as in a conventional Fermi liquid, albeit with slightly-modified numerical parameters (Nozières, 1974). Hence, for $k_B T$, $eV$, and $g\mu_B B$ $<< k_B T_K$ the Kondo spectral function and the associated transport through a spin-½ Kondo dot display quadratic dependence on external parameters, reflecting the $E^2$ scattering rate of quasiparticles in a Fermi liquid,

$$G(X) \sim G_0(1 - C(X/k_B T_K))^2 \qquad X = k_B T, eV, g\mu_B B << k_B T_K. \qquad (7)$$

$G_0 = G(X=0)$, and the constant $C$ is slightly different for the three perturbations.

In the opposite extreme, $k_B T$, $eV$, or $\mu_B B >> k_B T_K$, Kondo transport is well described by a perturbative treatment of the excitation and shows a logarithmic dependence on energy,

$$G(X) \sim 1/ln^2(X/k_B T_K) \qquad X = k_B T, eV, g\mu_B B >> k_B T_K. \qquad (8)$$

The two regimes evolve smoothly and monotonically into one another with no sharp features appearing at $T_K$. No analytic expression linking the two regimes has been derived, though an empirical expression (Goldhaber-Gordon et al., 1998b) that matches numerical renormalization group (NRG) calculations (Costi et al., 1994) over the entire temperature range has been widely adopted by experimentalists. We note that even though the three perturbations have similar effects on conductance, the mechanism is slightly different in each case, as discussed in the next two sections. The evolution of shot noise is predicted to follow forms similar to equations (7) and (8) (Meir and Golub, 2002), but with a universal average fractional charge of $5/3e$ (Sela et al., 2006).

### 2.3.3 Temperature Dependence

The conductance behavior predicted by equations (7) and (8) can be seen in temperature-dependent conductance measurements of a Kondo dot, as shown in Fig. 5. As temperature is raised, the conductance in the odd valleys decreases due to suppression of the Kondo effect. In contrast, conductance in the even valleys increases due to thermal broadening of the bare quantum dot levels (Pustilnik and Glazman, 2004). The scaled temperature dependent conductance for several gate voltage values in the Kondo valley is shown in Fig 5(b). The Kondo conductance shows the characteristic $T^2$ saturation at low temperatures, crossing over to a logarithmic temperature dependence as $T$ approaches $T_K$. The empirically derived formula alluded to above,

$$G(T) = G_0 \left( T_K'^2 / (T^2 + T_K'^2) \right)^s, \qquad (9)$$

provides a remarkably good fit over the whole temperature range (Goldhaber-Gordon et al., 1998b). The constant $s$ determines the slope of the conductance falloff. Using a value s=0.22 matches the slope found in numerical renormalization group (NRG) calculations for spin-½ Kondo systems (Costi et al., 1994). Here, $T_K' = T_K / (2^{1/s} - 1)^{1/2}$, which is equivalent to defining $T_K$ such that $G(T_K) = G_0/2$, as noted earlier. The extracted values of $T_K$ and $G_0$ across the Kondo valley are shown in Fig. 6. As predicted by equation (5), $T_K$ is maximal when the quantum level resides close to $E_F$.

At temperatures above $T_K$, non-Kondo conductance channels can develop (Pustilnik and Glazman, 2004), which will lead to deviations from equation (9). Some quantum dot devices exhibit parallel non-Kondo conduction channels, which are frequently accounted for by adding a temperature-independent offset to the Kondo form (equation (9)). Though convenient, the validity of using such an offset in analyzing Kondo behavior is not yet well-established.

### 2.3.4 Dependence on Bias and Magnetic Field

The $E^2$ dependence of the Kondo spectral function can be probed by applying a bias across the dot (Goldhaber-Gordon et al., 1998a, Cronenwett et al., 1998). The spectrum shows a narrow zero bias anomaly, referred to as the Kondo resonance, that is present across the whole odd-occupancy Coulomb valley (Fig. 7). For small biases, the Kondo conductance peak falls off with $V^2$. The Kondo peak broadens with increasing temperature, and the extrapolated zero temperature width is proportional to $T_K$ (van der Wiel et al., 2000). Exact interpretation of the Kondo peak shape is complicated by non-equilibrium processes (Wingreen and Meir, 1994, Schiller and Hershfield, 1995), which are discussed further in Sec. 3.2.

An applied magnetic field lifts the degeneracy between the two spin states on the quantum dot. The asymmetry between the two spin states acts to suppress Kondo correlations in much the same way as the asymmetry created by an applied bias. Progressively higher magnetic fields split the Kondo spectral function, which can be probed by measurement of differential conductance as a function of bias (Goldhaber-Gordon et al., 1998a, Cronenwett et al., 1998). These measurements are discussed in Sec. 3.2.

### 2.3.5 Universal Scaling

The similarity of the effects of temperature, bias, and magnetic field on the Kondo effect is neatly encompassed by a series of theoretically predicted universal scaling relations. In the context of Kondo transport, the behavior of different Kondo systems can be collapsed onto a single curve once each curve is scaled appropriately using only the Kondo temperature and zero temperature conductance, $T_K$ and $G_0$, respectively (Ralph et al., 1994, Schiller and Hershfield, 1995). The relation between temperature and bias can be expressed as

$$\frac{(G(V=0,T) - G(V,T))/G_0}{CT^{\alpha}} = F(eV/k_BT).$$ (10)

Here, $F(eV/k_BT)$ is a universal function that depends on the type of Kondo effect, but not on the exact details of the Kondo system at hand. The scaling constants $G_0$ and $C$ are defined by the low temperature expansion for $G(T)$ (equation (7)). The exponent $\alpha = 2$ for a spin-½ quantum dot, but can take on different values in more exotic Kondo systems (Ralph et al., 1994). Though the scaling properties of Kondo systems are frequently used in analyzing Kondo measurements (Fig. 5(b)), few experimental measurements of universal Kondo scaling functions or the expected deviation from universality (Majumdar et al., 1998), have been reported to date (Ralph et al., 1994).

## 2.4    Transmission Phase Shift and Unitary Limit

The zero-temperature Kondo conductance through a two lead quantum dot can also be derived via the Landauer formalism and Friedel sum rule, giving

$$G_0 = \frac{2e^2}{h} \left[ \frac{2\Gamma_1\Gamma_2}{\Gamma_1^2 + \Gamma_2^2} \right]^2 \frac{1}{2} \sum_{\sigma} \sin^2(\delta_{\sigma}),$$ (11)

where $\Gamma_i$ is the rate of tunneling between the local site and the i[th] lead, and $\delta_{\sigma}$ is the transmission phase shift of the σ spin channel (see Pustilnik and Glazman, 2004). For tunneling through a quantum dot in the Coulomb blockade regime, the transmission phase shift is changed by π each time one CB peak is tuned through the Fermi energy. Surprisingly, due to the equivalence of the two spin states brought about by the Kondo effect, the transmission phase shift evolves instead by π/2 in the Kondo regime at $T=0$ (Fig. 8(d)) (Gerland et al., 2000). As the temperature is raised to $T \sim T_K$ the phase shift evolves to the more conventional π phase shift, as seen in Fig. 8(b).

While the Kondo phase shift cannot be determined in bulk measurements, several mesoscopic measurements can determine it directly (Ji et al., 2000, Sato et al., 2005). One of the most successful approaches to measuring transmission phase shifts involves using an Aharonov-Bohm (AB) interferometer geometry like the one shown in Fig. 9(a) (Yacoby et al., 1995). Electrons can either tunnel through the quantum dot or travel ballistically through a reference arm. The relative phase of these two paths depends on the magnetic flux through the loop and the transmission phase shift acquired in tunneling through the quantum dot. The transmission phase shift is deduced by examining how the interference pattern changes as a function of the energy $\varepsilon_0$ of a localized state on the quantum dot. The Kondo phase shifts extracted from these experiments are somewhat larger than the theoretically predicted value, as seen in Fig. 9(c) (Ji et al., 2000, Ji et al., 2002). Transmission phase shifts can be measured using other two path interference geometries, such as a quantum wire side-coupled to a quantum dot (Sato et al., 2005). In these devices the measured phase shift is closer to the theoretically predicted value of π/2. More recent experiments (Avinun-Kalish et al., 2005, Neder et al., 2006) and theoretical work (Aharony et al., 2003, Aharony and Entin-Wohlman, 2005, Jerez et al., 2005) have elucidated some of the intricacies of phase shifts measured in mesoscopic electron interferometers.

An additional consequence of the equivalence of the two spin transport channels is that the zero-temperature conductance through a Kondo dot is $2e^2/h$ for symmetric reservoir-dot coupling (equation (11)). In contrast, the maximum conductance on a conventional Coulomb charging peak is only $e^2/h$ in the absence of Kondo correlations, since only one spin state can be transmitted at a time. The configuration of equal reservoir coupling and $T=B=V=0$ is commonly referred to as Kondo in the unitary limit. This limit can be achieved experimentally as long as $k_B T$, $eV$, $g\mu_B B << k_B T_K$ and the coupling asymmetry is small compared to $\Gamma$, as observed by van der Wiel et al. (van der Wiel et al., 2000).

# 3   Kondo Effect in Modified Environments

In the previous section we briefly discussed how the Kondo state is suppressed by external perturbations such as temperature and magnetic field.  In this section we provide a more general description of how the spin-½ Kondo state evolves in response to modified environments. In addition to temperature, magnetic field, and bias, non-ideal conduction reservoirs and other external couplings can affect Kondo behavior. The effects of perturbations on the Kondo state can be categorized as follows: (1) the Kondo state is suppressed due to decoherence of the Kondo singlet (e.g. suppression of Kondo by thermal fluctuations); (2) a non-Kondo ground state becomes the lowest energy ground state (e.g. formation of spin-polarized coupling ground state in high magnetic field); or (3) a new Kondo state is created. The Kondo energy scale ($k_B T_K$) is useful scale for determining how a perturbation will affect the Kondo state. Weak perturbations ($E < k_B T_K$) typically only partially suppress the Kondo state, while stronger perturbations ($E > k_B T_K$) are needed to significantly alter or destroy the Kondo singlet.

## 3.1   Decoherence of the Kondo State

We first discuss how various processes can lead to decoherence of the Kondo singlet. The Kondo singlet state is created by spin-flip co-tunneling processes between conduction electrons and the local state.  Non-Kondo processes, such as emission or absorption of a phonon or photon, disturb the necessary superposition of electronic states in the Kondo singlet. At zero temperature and bias, the phase space available for these non-Kondo processes is nearly eliminated and delocalized electrons within length $\xi_K$ (equation (7)) form a Kondo singlet with the local spin. At finite temperature, thermal fluctuations drive non-Kondo transitions, progressively destroying these correlations. These non-Kondo transitions are facilitated by smearing of the sharp Fermi distribution of the leads with temperature, which increases the available phase space for the transitions.  Similarly, a small bias ($eV < k_B T_K$) applied across the dot also opens up phase space for non-Kondo transitions, leading to a suppression of the Kondo state. Larger biases have a more profound affect on the Kondo state, as discussed later in this section.

Coupling the quantum dot to other noise sources can have similar decohering effects. M. Avinun-Kalish et al. experimentally investigated the suppression of the Kondo state due to a capacitively coupled QPC (Avinun-Kalish et al., 2004). Conceptually, the QPC can be treated either as a source of shot noise or as a quantum dot charge measurement device, as discussed in Sec. 2.3.1. The authors observed stronger-than-expected suppression of Kondo transport, which they attributed to the long residence time of an electron in the Kondo cloud that extends well into the leads (Avinun-Kalish et al., 2004). The Kondo state can also be decohered by irradiating the sample with microwave radiation (Kaminski et al., 1999, Elzerman et al., 2000). Here the random population and decay of excited states by photons takes over the role of thermal fluctuations in destroying the Kondo state.

## 3.2   Non-Equilibrium Kondo Effect

An applied bias across the quantum dot can have several interesting consequences on the Kondo state. For a quantum dot coupled equally to two leads at the same potential, the Kondo screening cloud spans both leads. An applied bias between the leads exposes the quantum dot to different Fermi levels, which opens up inelastic (decohering) channels for electron scattering between different leads. Kondo exchange with each individual lead, however, is not affected. For biases above $k_B T_K$, the Kondo spectral function splits into two resonances centered on the Fermi energies of each reservoir, as shown in Fig. 10(a) (Wingreen and Meir, 1994, Schiller and Hershfield, 1995). The two Kondo resonances slowly fade with increasing bias due to decoherence associated with the exchange of electrons between the leads, but they remain visible for biases far exceeding $k_B T_K$. Two-terminal quantum dot measurements cannot accurately map out the non-equilibrium spectral function, but instead measure a convolution of two evolving spectral functions. The splitting can be observed experimentally through the addition of a weakly coupled third lead (Lebanon and Schiller, 2001, Leturcq et al., 2005), as shown in Fig. 10(b). An alternative approach to probing the out-of-equilibrium Kondo spectral function has been investigated using a split Fermi distribution in a quantum wire side-coupled to a quantum dot (De Franceschi et al., 2002).

### 3.2.1 Recovery of Kondo State

An applied bias can also provide the necessary energy to form a Kondo singlet when the Kondo ground state is not otherwise the favored configuration. The first example of this phenomenon is the recovery of the Kondo effect at finite bias in an applied magnetic field. A magnetic field breaks the spin degeneracy of the quantum dot ground state favoring a spin-polarized ground state over the Kondo singlet. As a result, spin-flip transitions incur an energy cost, leading to a characteristic suppression of Kondo conductance, as described in Sec. 2.3.2. The spin-flip energy cost can be compensated by applying a bias across the quantum dot. Here, the Kondo state re-emerges when the applied bias is equal to the spin-flip energy ($eV \sim g\mu_B B$) (Goldhaber-Gordon et al., 1998a, Cronenwett et al., 1998, Kogan et al., 2004a). Experimentally, this produces two Kondo peaks, separated by twice the spin-flip energy (Fig. 11). Surprisingly, the measured splitting of the Kondo resonance exceeds the expected Zeeman splitting, an effect which is not fully understood (Kogan et al., 2004a).

High frequency microwave radiation ($hf > k_B T_K$) can break up the Kondo singlet and destroy the Kondo effect, as described previously. On the other hand, microwave radiation can also drive Tien-Gordon photon-assisted tunneling processes (Kouwenhoven et al., 1994). By tuning the applied bias to the photon energy ($eV \sim hf$), a unique Kondo state can exist in which coherent spin-flip tunneling events coincide with simultaneous absorption or emission of photons. Kogan and co-workers found that such a state can be achieved experimentally if the monochromatic microwave radiation intensity is tuned so that $eV_{osc} \sim hf$ (Kogan et al., 2004b). These conditions ensure that the photon field is strong enough to modify Kondo processes, but not strong enough to decohere them altogether. The experimental signature of this state is a splitting of the Kondo peak into two peaks separated by twice the photon frequency, as seen in Fig. 12.

### 3.2.2 Cross-over to Co-tunneling

In the examples of non-equilibrium Kondo effects presented here, finite bias is essential to the creation of the Kondo state even though it also introduces decoherence processes. Inevitably, the Kondo state will decohere completely when exposed to strong enough perturbations ($E >> k_B T_K$). Incoherent, or inelastic, co-tunneling processes can still enhance conductance in this regime, though to a much lesser extent than the analogous coherent processes that compose the Kondo effect (De Franceschi et al., 2001). The transition from coherent Kondo tunneling to incoherent co-tunneling has been investigated by several groups (Sasaki et al., 2000, Kogan et al., 2004a). Figure 11(c,d) shows the seamless transition from Kondo to incoherent co-tunneling as a function of an applied magnetic field (Kogan et al., 2004a).

## 3.3 Kondo in Landau Levels

Though an applied magnetic field generally suppresses the Kondo state, several novel Kondo states can form at intermediate fields. The electronic states of a 2DEG in magnetic fields are Landau levels (LL), which for our purposes are circular orbits at quantized multiples of the cyclotron energy. Quantum dot states can acquire LL character once the electron cyclotron radius is comparable to the spatial dimensions of the dot. The quantum dot LL states behave in much the same way as regular quantum dot states and can exhibit Kondo behavior (Schmid et al., 2000, Sprinzak et al., 2002, Stopa et al., 2003, Keller et al., 2001). Typically, the lowest spin-degenerate LL is most strongly coupled to the leads, and Kondo processes occur when this level contains an unpaired spin (Stopa et al., 2003, Keller et al., 2001). The occupancy of the LL can be tuned by changing the total number of electrons with an applied gate voltage or by redistributing the existing electrons among the LLs using an applied magnetic field. As a result, the Kondo conductance shows a "checkerboard" pattern as a function of gate voltage and magnetic field, as shown in Fig. 13 (Schmid et al., 2000, Sprinzak et al., 2002, Stopa et al., 2003, Keller et al., 2001). At higher magnetic fields, the Zeeman splitting spin-polarizes the leads, and the Kondo conductance vanishes. In high magnetic fields electrons can form closed orbits around a depleted region of a 2DEG. The resulting "antidot" can act as a magnetic impurity and exhibit the Kondo effect (Kataoka et al., 2002).

## 3.4 Modified Conduction Reservoir

In order for Kondo correlations to persist, the quantum dot must be able to freely exchange electrons with the leads. Metallic leads facilitate electron exchange by providing a large smooth density of states near the Fermi level. However, distinctive Kondo states can exist even if the leads have additional electronic structure.

### 3.4.1 Finite level spacing (Kondo Box)

A natural question to ask is what happens to Kondo processes when the mean level spacing of the leads ($\Delta$) becomes comparable to the Kondo temperature ($T_K$) (Thimm et al., 1999, Simon and Affleck, 2002). Thimm et al. (1999) explored this question theoretically in the context of a single electron transistor and found three unique signatures: (1) the Kondo resonance splits up into a series of subpeaks, (2) the conductance depends on the even/odd occupancy of the combined lead-quantum dot system, and (3) quantum dot transport exhibits Fano-like line shapes with anomalous temperature dependence. A "Kondo Box" exhibiting these features has yet to be quantitatively studied experimentally despite several interesting approaches (Odom et al., 2000, Manoharan et al., 2000, Booth et al., 2005).

### 3.4.2 Magnetic Leads

Ferromagnetic leads create spin imbalances, which affect the dynamics of Kondo spin flip exchange. The behavior of these systems depends on the relative magnetic orientation of the two leads connected to the quantum dot (Sergueev et al., 2002, Martinek et al., 2003, Choi et al., 2004). If both leads have the same magnetic orientation, the reservoir spin imbalance acts similarly to a magnetic field. Here, the Kondo spectral function is suppressed and split, whenever the quantum dot is not electron-hole symmetric ($\varepsilon_0 \neq -U/2$) (Choi et al., 2004). The Kondo effect for anti-parallel magnetic lead alignment is equivalent to the non-magnetic case, though with a different non-equilibrium behavior. Experimental realizations of ferromagnetic coupled leads in mesoscopic systems have been hampered by the difficulty of reliably creating magnetic semiconductor structures. Some of the predicted magnetic behavior has been seen, however, in single molecular transistor (Pasupathy et al., 2004).

### 3.4.3 Superconducting and Luttinger-Liquid Leads

More exotic systems have also been considered, such as coupling a quantum dot to superconducting (Fazio and Raimondi, 1998, Sun et al., 2001) or Luttinger-Liquid leads (Kane and Fisher, 1992, Lee and Toner, 1992, Furusaki and Nagaosa, 1994). In both cases, the leads have no density of single electron states at the Fermi level, where the Kondo screening cloud would normally form. Therefore, the Kondo ground state competes with the native ground state of the leads. For superconducting leads, the Kondo effect will develop if the Kondo energy scale, $k_B T_K$, exceeds the superconducting single-electron excitation energy gap, $\Delta$, as seen experimentally in carbon nanotube quantum dots (Buitelaar et al., 2002). Kondo behavior in Luttinger-liquids depends sensitively on the Luttinger parameter of the leads (Komnik and Gogolin, 2003), but has not been examined in mesoscopic experiments.

# 4   Exotic Kondo Systems

The Kondo effect, as described by the Anderson model (equation (1)), arises when a localized degenerate state is coupled to a reservoir of conduction electrons. Though our discussion has focused on spin-½ degeneracy, the Kondo effect can also arise from a two-fold degeneracy with other physical origins (Fig. 14), or from a higher local degeneracy.  In this section we review recent mesoscopic systems that demonstrate such scenarios.

## 4.1   Non-spin Kondo systems

### 4.1.1   Two-fold electrostatic degeneracy

One of the most striking examples of non spin-½ Kondo system is the spinless Kondo effect observed in a capacitively coupled double quantum dot system (Wilhelm et al., 2002).  The experimental configuration (shown in Fig. 15(a)), consists of two parallel quantum dots, each coupled to its own independent set of leads. The two dot-lead systems cannot exchange electrons with each other, but can communicate via electrostatic coupling between the two quantum dots. A charge degeneracy can be created by tuning the two quantum dots so that the quantum state corresponding to having an extra electron on dot 1 $(N_1+1, N_2)$ is degenerate with the state corresponding to an extra electron on dot 2 $(N_1, N_2+1)$. This charge degeneracy takes over the role normally played by spin degeneracy, mapping this system onto the Anderson impurity model (Pohjola et al., 2000, Wilhelm et al., 2001).

Transport through one of the dots as a function of the occupancy of both dots reveals a honeycomb pattern (Fig. 15(b)), similar to that seen in transport through double quantum dot systems (van der Wiel et al., 2003). The faint conductance feature on side b of the hexagon in Fig. 15(b) (at the charge degeneracy point described earlier) is the feature of interest in this experiment. While single electron transport is forbidden due to Coulomb blockade, two-electron processes give rise to non-zero conductance. Here an electron jumps off one dot while another electron simultaneously jumps onto the other. These processes drive the Kondo effect, screening the electrostatic (as opposed to magnetic) degeneracy, and giving rise to a non-zero conductance.

### 4.1.2   Two-fold orbital degeneracy

A more common non-spin degeneracy in quantum dot systems is orbital degeneracy, which typically arises from spatial symmetries. The orbital Kondo effect has been observed in a circular quantum dot (Sasaki et al., 2004). Carbon nanotubes have an intrinsic orbital degeneracy due to their structure, and several groups have reported Kondo effects stemming from this degeneracy (Nygård et al., 2000, Jarillo-Herrero et al., 2005b, Jarillo-Herrero et al., 2005a). Non-degenerate orbital states can be brought into degeneracy by applying a magnetic field, as shown in Fig. 14(b), leading to analogous Kondo effects (Sasaki et al., 2004). Spin degeneracy may also combine with orbital degeneracy, leading to exotic Kondo physics as described in the following subsections.

## 4.2 Multiple Degeneracy Kondo systems

Quantum dots with a multiple degenerate grounds state can exhibit a variety of interesting Kondo effects. The Kondo behavior of these systems depends on the allowed transitions between the N-degenerate states. If transitions can occur equally between any pair of degenerate states, the system has SU(N) symmetry and is described by the Coqblin-Schrieffer model (Coqblin and Schrieffer, 1969). The Kondo temperature is predicted theoretically to be larger in these systems than in equivalent spin-½ Kondo systems, and depends exponentially on N:

$$T_K \sim D e^{-\frac{1}{N\rho(E_F)J}},$$  (12)

where $J$, $\rho(E_F)$, and $D$ are the exchange coupling, density of states, and interacting electron bandwidth, respectively. Like the spin-½ Kondo ground state, the SU(N) Kondo ground state is no longer degenerate. In contrast, for an impurity with a high spin degeneracy ($S > ½$), a single conduction channel will only couple states with $\Delta S_z = 1$. In these systems, the Kondo effect will underscreen the magnetic moment, leaving behind a ground state of lower degeneracy, associated with a net spin reduced by ½ from its native value (Mattis, 1967). Other systems, such as the singlet-triplet degeneracy discussed later, can have a more complicated coupling structure (Eto, 2005).

### 4.2.1 Spin and orbital degeneracy: SU(4) Kondo Effect

When the electronic state of a system contains both orbital and spin degeneracy, the same delocalized lead electrons that magnetically screen the unpaired spin may also electrostatically screen the orbital degeneracy. The two degrees of freedom can produce an SU(4) Kondo effect, which has been observed in vertical GaAs dots (Sasaki et al., 2004) and carbon nanotubes (Jarillo-Herrero et al., 2005b). In the vertical quantum dot system, $T_K$ is not large enough for the spin-½ Kondo effect to be observed in the absence of orbital degeneracy (odd valleys in Fig. 16(a)). Kondo conductance enhancement is observed, however, when the dot is tuned to an orbital crossing point using an applied magnetic field. Here an SU(4) four-fold degeneracy is present, which enhances the Kondo temperature (Eto, 2005). This "doublet-doublet" Kondo state produces a zero bias Kondo peak, as seen in Fig. 16(b).

Carbon nanotubes have intrinsic orbital degeneracy in addition to spin degeneracy, which can lead to SU(4) Kondo physics (Jarillo-Herrero et al., 2005b). Due to electron-hole symmetry, the SU(4) Kondo effect is observed when the two orbitals are occupied by either one or three electrons (Fig. 17(a)). The four-fold degeneracy responsible for the Kondo ground state is demonstrated by the splitting of the Kondo resonance into two peaks with magnetic field (Fig. 17(b)). The magnetic field suppresses the SU(4) Kondo effect by first lifting the orbital degeneracy. The resulting spin-½ Kondo effect survives until the Zeeman splitting becomes comparable to the Kondo temperature. A pure orbital SU(2) Kondo effect is attained at orbital level-crossings, which one can attain by applying magnetic field (dashed lines in Fig. 17(a)). In these experiments the Kondo temperature for the SU(4) Kondo effect is larger than the Kondo temperature for the SU(2) Kondo effect, as predicted by equation (12). It is not yet clear why the electrons in the (metal) leads have the requisite four-fold symmetry to fully screen the local state on the nanotube.

### 4.2.2  Triplet and Singlet-Triplet Kondo Effect

The ground state of a quantum dot with an even number of electrons is usually a spin singlet at zero magnetic field, where each occupied orbital contains a pair of electrons. The triplet state will have lower energy, however, if the exchange energy gained for parallel spin filling exceeds the level separation between adjacent orbitals (Tarucha et al., 2000). A quantum dot with S=1 no longer has the same spin symmetry as the spin-½ reservoir electrons. As discussed earlier, Kondo correlations will partially screen the S=1 quantum dot, leaving behind a residual spin-½, as long as only a single reservoir mode is coupled to the quantum dot. Perhaps due to a low $T_K$ predicted for these systems (Wan et al., 1995, Izumida et al., 1998), a pure spin-1 Kondo effect has not been unambiguously observed in mesoscopic systems.

Alternatively, singlet and triplet state can be brought into degeneracy using an applied magnetic field. At low magnetic fields the Zeeman splitting of the triplet is negligible and the singlet-triplet degeneracy point is four-fold degenerate, though not SU(4) symmetric (Eto and Nazarov, 2000, Pustilnik and Glazman, 2000). The singlet-triplet Kondo temperature is expected to be higher than for spin-½ Kondo systems and comparable to that found in SU(4) Kondo systems (Eto, 2005). A singlet-triplet Kondo effect has been observed in transport measurements through vertical quantum dots (Sasaki et al., 2000, Sasaki et al., 2004)  (shown in Fig. 16)), lateral quantum dots (Schmid et al., 2000, van der Wiel et al., 2002, Kogan et al., 2003), and carbon nanotubes (Nygård et al., 2000, Jarillo-Herrero et al., 2005b, Jarillo-Herrero et al., 2005a)  (Fig. 17(a)). Even if the singlet and triplet states are not energetically degenerate a finite bias can compensate for the energy difference and lead to a non-equilibrium singlet-triplet Kondo Effect, as seen recently in a carbon nanotube quantum dot (Paaske et al., 2006).

### 4.2.3  Two-stage Kondo Effect

A system on the border of a singlet-triplet transition can exhibit a remarkable two-stage Kondo effect. If coupled to two reservoir modes, the transport properties of a $S$=1 quantum dot are theoretically predicted to have a non-monotonic dependence on temperature and bias (Pustilnik and Glazman, 2001). The two modes create two different Kondo screening channels with different associated Kondo temperatures, $T_{K1}$ and $T_{K2}$. At unitary coupling, the two modes interfere and reduce Kondo conductance when T is below both $T_{K1}$ and $T_{K2}$. A second screening channel can form in systems containing only one tunneling mode if the dot-lead couplings are asymmetric (Pustilnik and Glazman, 2001). A quantum dot just on the singlet side of the singlet-triplet transition is also predicted to undergo a similar two stage Kondo effect, even if only a single tunneling mode is coupled to the dot (Hofstetter and Schoeller, 2002). Such two-stage Kondo effects have been observed in lateral quantum dots (van der Wiel et al., 2002, Granger et al., 2005).  In experiments by van der Wiel, a magnetic field tunes the quantum dot near a singlet-triplet degeneracy point. Conductance measurements show a dip in the zero-bias anomaly at low temperatures (Fig. 18). From the available data, it is not possible to unambiguously distinguish whether a singlet or triplet ground state is responsible for the observed effect.

# 5 Recent Developments and Future Directions

As we have seen, the designability and in-situ control of parameters in semiconductor quantum dots has enabled highly-quantitative tests of Kondo physics, and observation of Kondo effects in new regimes and scenarios. This new approach has brought fresh insight and excitement to what was legitimately considered one of the best understood phenomena in solid state physics. The continuing flow of theoretical predictions and experimental investigations – around 100 papers per year in mesoscopic physics alone – attests that Kondo physics remains a rich and vibrant field. We conclude our review with a few remarks about the future of mesoscopic Kondo experiments, as suggested by recent trends.

## 5.1 Spatial and Temporal Kondo Physics

Mesoscopic experiments have illuminated the effect on Kondo transport of external parameters (e.g. temperature, magnetic field) and dot-lead coupling.  In short, spin-½ Kondo phenomenology is broadly well-understood. However, several important issues remain on the cutting edge: How do a pair of magnetic impurities interact (Jeong et al., 2001, Craig et al., 2004, Simon et al., 2005, Vavilov and Glazman, 2005)? What are the dynamics and spatial extent of the Kondo screening cloud (Sorensen and Affleck, 1996, Nordlander et al., 1999) and how do these depend on the dimensionality of the screening reservoirs (Simon and Affleck, 2002)? How can we understand and control the processes that lead to decoherence of the Kondo state, especially out of equilibrium (Avinun-Kalish et al., 2004, Kogan et al., 2004a, Leturcq et al., 2005)?

## 5.2 Novel Kondo Systems

As discussed in Sections 3 and 4, design of a quantum dot's states has enabled studies of a broad family of Kondo effects with different local degeneracies. However, several proposed exotic Kondo systems are not yet addressed by extensive experiments. Controlling the electronic properties of the conduction reservoir, as discussed in Sec. 3.4, has proven difficult in mesoscopic systems. Manipulating the coupling of quantum dots to superconducting, magnetic, or one-dimensional leads will be exciting, and is likely to advance fastest in hybrid molecular-metal systems, in which the material of the metal leads can be selected to yield desired electronic structure (Buitelaar et al., 2002, Pasupathy et al., 2004). Tantalizing theoretical predictions abound for multi-channel Kondo systems, such as the two-channel Kondo effect (Cox and Zawadowski, 1998, Matveev, 1995, Oreg and Goldhaber-Gordon, 2003, Pustilnik et al., 2004), which occurs when two reservoirs independently attempt to screen a spin degenerate state. The resulting frustrated system is predicted to exhibit both local non-Fermi Liquid behavior and a model quantum phase transition, as recently observed by some of the present authors in a novel double-dot geometry (Potok et al., 2006).  This offers hope that the tools of mesoscopic physics can bear on complex correlated electrons systems normally realized only in bulk systems.  Finally, several experimental observations indicate that the celebrated 0.7 structure seen in QPC transport might have Kondo origins (Cronenwett et al., 2002).  This effect is yet to be fully understood and should continue to garner both experimental and theoretical attention.

## Figure Captions

*Figure 1*

The Kondo effect has been seen in a variety of quantum dot geometries such as **(a)** lateral quantum dots (SEM image), **(b)** vertical quantum dots (schematic) (adapted with permission from Macmillan Publishers Ltd: Sasaki et al., Nature 405, 764 (2000)), and **(c)** carbon nanotubes (SEM and schematic) (adapted with permission from Macmillan Publishers Ltd: Nygård et al., Nature 408, 342 (2000)).

*Figure 2*

Schematic diagram of the Anderson model as applied to a singly occupied spin-degenerate quantum dot state coupled to a single electron reservoir. The parameters are discussed in the text.

*Figure 3*

A magnetic impurity in a sea of conduction electrons. **(a)** Above the Kondo temperature ($T_K$) the conduction electrons are only weakly scattered by the magnetic impurity. **(b)** Below $T_K$ the conduction electrons screen the local spin to form a spin-singlet. The formation of the Kondo screening cloud enhances the effective scattering cross-section of the magnetic impurity.

*Figure 4*

Linear conductance through the lateral quantum dot shown in Fig. 1(a), as a function of gate voltage. From left to right, the quantum dot electron occupancy increases by four and three in (a) and (b), respectively. **(a)** Kondo regime: For strong lead-dot coupling, the Kondo effect enhances conductance for odd occupancy at 90 mK. At higher temperatures (800 mK) the Kondo effect is partially suppressed and transport proceeds through temperature broadened single-particle levels. **(b)** Coulomb blockade (CB) regime: For weak dot-lead coupling ($\Gamma \ll U, \Delta$), the Kondo temperature is lower than the base temperature of the experiment (90 mK) and transport occurs only at the charge degeneracy points. Note: the peaks in (b) have been shifted by +120mV and do not correspond to the same electronic states as in (a). Reprinted with permission from [Goldhaber-Gordon et al. Phys. Rev. Lett. 81, 5225 (1998)]. Copyright (1998) by the American Physical Society.

*Figure 5*

**(a)** Temperature dependence of the conductance in the Kondo valley. The vertical dashed lines indicate the location of the charge degeneracy points (i.e. $\varepsilon_0 = 0$ or $\varepsilon_0 + U = 0$), as deduced from data for $T \gg T_K$. Inset: The extrapolated linear temperature dependence of the Coulomb blockade peak width at $T=0$ gives $\Gamma = 295 \pm 20$ μeV. The slope of the line determines the conversion factor between the applied gate voltage ($V_g$) and $\varepsilon_0$. **(b)** Normalized conductance ($G/G_0$) as a function of normalized temperature ($T/T_K$) for several points along the Kondo valley. Here $\widetilde{\varepsilon} = \varepsilon_0 / \Gamma$. $G_0$ and $T_K$ were extracted by fitting the temperature dependence to equation (9). The data shown here are from the lateral quantum dot device shown in Fig. 1(a). Reprinted with permission from [Goldhaber-Gordon et al. Phys. Rev. Lett. 81, 5225 (1998)]. Copyright (1998) by the American Physical Society.

### Figure 6

**(a)** Values of $T_K$ across the Kondo valley shown in Fig. 5, extracted by fitting the temperature dependent conductance to the Kondo form (equation (9)). The behavior is well described by equation (5) (solid line). Inset: Expanded view of the left side of the figure, showing the quality of the fit. **(b)** Values of $G_0$ across the Kondo valley extracted from the temperature fits in part (a) (open circles) or from base temperature conductance (crosses). The solid line shows $G_0(\varepsilon_0)$ predicted by Wingreen and Meir (1994). $G_{max} = 0.49 \ e^2/h$ and $0.37 \ e^2/h$ for the left and right peak, respectively. Reprinted with permission from [Goldhaber-Gordon et al. Phys. Rev. Lett. 81, 5225 (1998)]. Copyright (1998) by the American Physical Society.

### Figure 7

**(a)** Low temperature (T = 12 mK) differential conductance (dI/dV) through a lateral quantum dot, as a function of gate voltage and applied bias in the Kondo regime. In the absence of the Kondo effect, dI/dV is non-zero only when a quantum dot energy level is aligned with the Fermi energy of one of the leads (*cf.* Kouwenhoven et al. (2003)). The resulting "Coulomb diamonds" are modified by the Kondo effect: a Kondo resonance appears at zero-bias across the whole Coulomb valley for odd occupancy. The Kondo effect also alters finite bias conduction through the quantum dot levels. **(b)** Temperature dependence of the Kondo resonance in the middle of the Kondo valley ($V_{gate}$ = -203.5 mV).

### Figure 8

Theoretical temperature and coupling dependence of the transmission amplitude ($|t_{d\sigma}|$) and phase ($\Phi_{d\sigma}$) for transport through a quantum dot, as a function of the normalized energy of the quantum dot level ($\varepsilon_0 / U$). Each pane plots results for three values of dot-lead coupling ($\Gamma$). The transmission amplitude is normalized to $2e^2/h$. **(a, b)** The left two panels show calculations for $T = \Gamma > T_K$. In this regime, transport occurs mainly through thermally broadened quantum dot levels. **(c, d)** The right two panels show calculations for $T$=0. The Kondo effect is in the unitary limit, giving a $2e^2/h$ conductance throughout the Kondo valley. In this regime, the phase shift is $\pi/2$ in the Kondo valley. Reprinted with permission from [Gerland et al. Phys. Rev. Lett. 84, 3710 (2000)]. Copyright (2000) by the American Physical Society.

### Figure 9

**(a)** SEM image of a mesoscopic Aharonov-Bohm interferometer with **(b)** rough schematic of the device. **(c)** Evolution of the transmission phase and interference visibility as a function of the quantum dot energy level in the Kondo regime. The phase shift plateaus at $\pi$ in the middle of the Kondo valley before rising to $1.5\pi$ when the quantum level is doubly occupied. **(d)** In the Coulomb blockade regime ($T_K < T_{base}$), the phase shift increases and then falls by $\pi$ each time the quantum dot occupancy is increased by one electron. Adapted with permission from [Li et al., Science 290, 779 (2000)]. Copyright (2000) AAAS.

*Figure 10*

**(a)** A schematic of the splitting of the Kondo spectral function with an applied bias. The relative amplitude of the Kondo resonance in each lead depends on the respective dot-lead coupling. In the extreme case, where one lead is well-coupled and the other is weakly-coupled (typical of STM experiments), the weakly-coupled lead can be treated as a spectroscopic probe (Lebanon and Schiller, 2001, Nagaoka et al., 2002). **(b)** The splitting of the Kondo resonance with applied bias can be probed by weakly coupling an additional lead. The plot shows the current through the weakly coupled lead as a function of the voltage difference ($\Delta V$) between the two strongly coupled leads. A Kondo conductance enhancement is observed when the Fermi energy of the weakly coupled lead is aligned with the Fermi energy of either one of the strongly coupled leads, leading to the cross-pattern seen in the figure. Adapted with permission from [Lecturcq et al. Phys. Rev. Lett. 95, 126603 (2005)]. Copyright (2005) by the American Physical Society.

*Figure 11*

Evolution of the Kondo resonance and the inelastic spin-flip co-tunneling gap with applied magnetic field for two different quantum dot devices. **Device 1: (a)** The solid line (dots) shows the evolution of the Kondo peak (co-tunneling gap) with applied field. **(b)** The Kondo peak splitting ($2\Delta_K$, solid squares) is larger than the spin-flip co-tunneling gap ($2\Delta$, open circles), indicating that $\Delta_K > g\mu_B B$. **Device 2: (c)** The split Kondo peak seamlessly evolves into an inelastic spin-flip co-tunneling signal as the magnetic field is increased. **(d)** The transition to inelastic co-tunneling is marked by a decrease in the measured gap, indicating again that $\Delta_K > g\mu_B B$. Reprinted with permission from [Kogan et al. Phys. Rev. Lett. 93, 166602 (2004)]. Copyright (2004) by the American Physical Society.

*Figure 12*

Splitting of the Kondo peak as a function of microwave irradiation amplitude. Reprinted with permission from [Kogan et al., Science 304, 1293 (2004)]. Copyright (2004) AAAS.

*Figure 13*

In a high magnetic field applied perpendicular to the quantum dot, the states of a quantum dot organize into Landau levels. If the dot is strongly coupled to its leads, the Kondo effect occurs whenever the occupancy of the outer Landau level (LL) is odd. As a result, the differential conductance shows a checkerboard pattern as the LL occupancy can be tuned by both a gate voltage and the magnetic field. Reprinted with permission from [Schmid et al. Phys. Rev. Lett. 84, 5824 (2000)]. Copyright (2000) by the American Physical Society.

*Figure 14*

**(a)** The spin-½ Kondo effect stems from the two-fold spin degeneracy of a quantum dot level. Degeneracies with origins other than spin can also lead to a completely equivalent Kondo effect. Such degeneracies can arise from: (i) two spatially separated sites of which only one can be occupied due to Coulomb repulsion, (ii) a spatial symmetry that leads to two degenerate orbitals, and (iii) the lattice symmetry of a carbon nanotube which gives rise to a twofold orbital degeneracy. Reprinted with permission from Macmillan

Publishers Ltd: R. M. Potok et al. Nature 434, 451 (2005). **(b)** Orbitals with different energies can be made degenerate by tuning their energies with an applied magnetic field. Shown here is the evolution of circular quantum dot states with magnetic field. Degeneracies occur at level crossings, such as the one indicated by the circle. Reprinted with permission from [Sasaki et al. Phys. Rev. Lett. 93, 017205 (2004)]. Copyright (2004) by the American Physical Society.

*Figure 15*

**(a)** Schematic of capacitively coupled double quantum dot system used to observe the spinless Kondo effect (scenario (i) in Fig. 14(a)). **(b)** Current through the upper quantum dot as a function of $V_{G1}$ and $V_{u,d}$ shows a honeycomb pattern due to electrostatic interaction with the lower quantum dot ($V_{DS,u} = 80$ μV; 'white' = no current). The spinless Kondo effect leads to the faint conductance observed along the line marked "b". Reprinted with permission from [Wilhelm et al., Physica E 14, 385 (2002)]. Copyright (2002) by Elsevier.

*Figure 16*

**(a)** Zero bias conductance measured in a vertical quantum dot as a function of gate voltage and magnetic field. Orbital states are brought into and out of degeneracy by the magnetic field, which leads to two types of Kondo effects. For odd occupancy, a doublet-doublet (D-D) Kondo effect arises at orbital crossings where a four-fold degeneracy is present. A singlet-triplet (S-T) Kondo effect occurs at level crossings for even occupancy. To the right of the dotted line, all electrons occupy the lowest Landau level. **(b)** The temperature dependence of the Kondo resonance for the S-T and D-D Kondo effects from 60 mK (thick solid line) to 1.5 K (dashed line). Both systems show a similar Kondo temperature. Reprinted with permission from [Sasaki et al. Phys. Rev. Lett. 93, 017205 (2004)]. Copyright (2004) by the American Physical Society.

*Figure 17*

**(a)** Conductance through a carbon nanotube quantum dot (shown schematically in the inset) as a function of magnetic field ($B$) and gate voltage ($V_G$). The solid lines highlight the evolution of the Coulomb blockade peaks with B. The Roman numerals indicate the number of electrons occupying the partially-occupied four electron shell at B=0, while the Arabic numbers represent the spin of the ground state. At $B$=0, an SU(4) Kondo effect occurs for single (I) and triple occupancy (III), while a singlet-triplet Kondo effect is observed for double occupancy (II). A purely orbital Kondo effect is observed at magnetic field-induced level crossings, as indicated by the horizontal dotted lines. **(b)** A magnetic field will split the four degenerate states (two spin ⊗ two orbital) that generate the SU(4) Kondo effect. Since the orbital magnetic moment is approximately three times larger than the spin magnetic moment in a carbon nanotube, the orbital states are split at a different rate than the spin states, and four peaks are observed at finite B-field. Reprinted with permission from Macmillan Publishers Ltd: Jarillo-Herrero et al., Nature 434, 484 (2005).

*Figure 18*

The two-stage Kondo effect is characterized by enhanced conductance at intermediate energies and suppressed conductance at low energies, as demonstrated by the dip in the zero-bias Kondo resonance.  Shown here is the temperature dependence of the Kondo peak from 15 mK (solid line) to 800 mK.  Reprinted with permission from [van der Wiel et al. Phys. Rev. Lett. 88, 126803 (2002)]. Copyright (2002) by the American Physical Society.

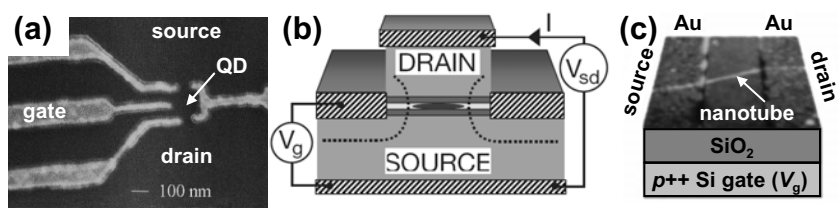

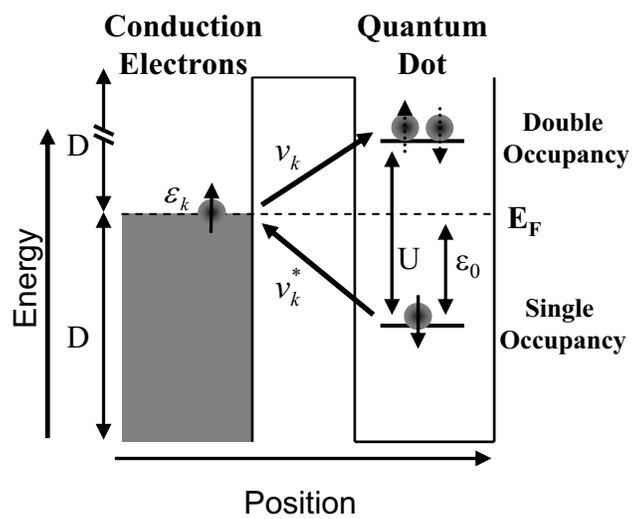

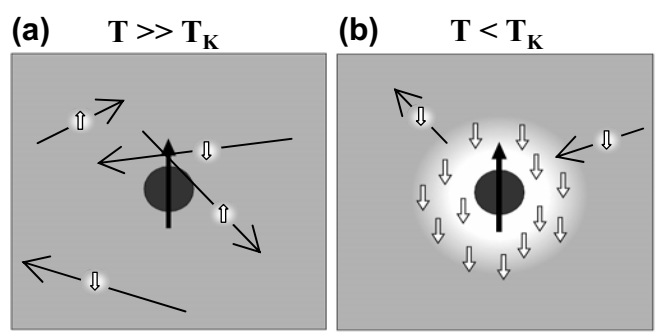

**(a)** T >> T$_K$    **(b)** T < T$_K$

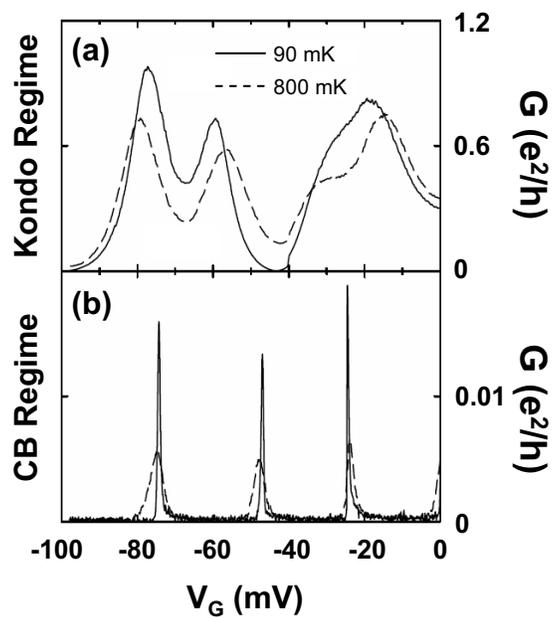

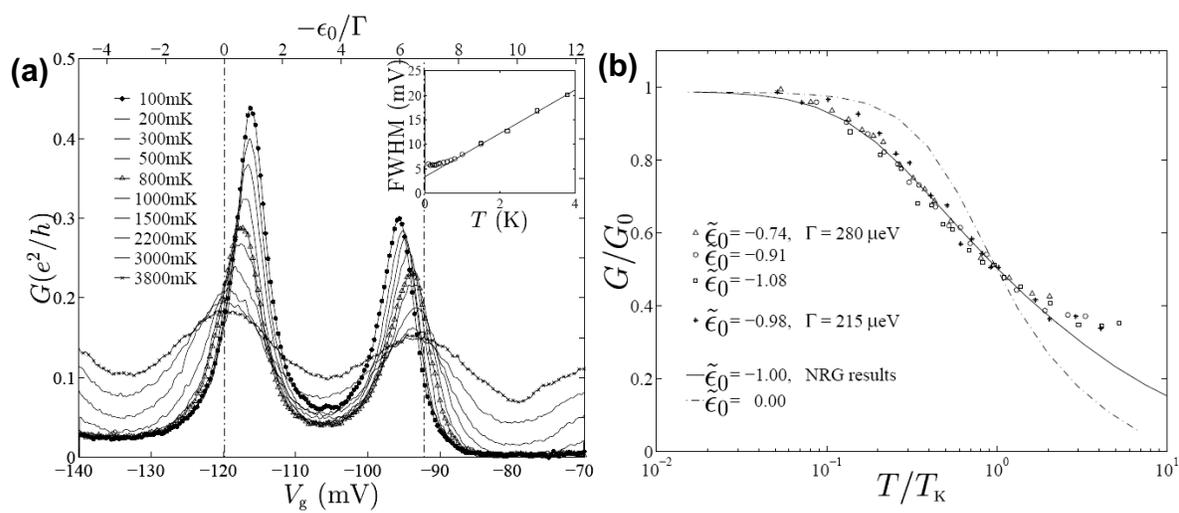

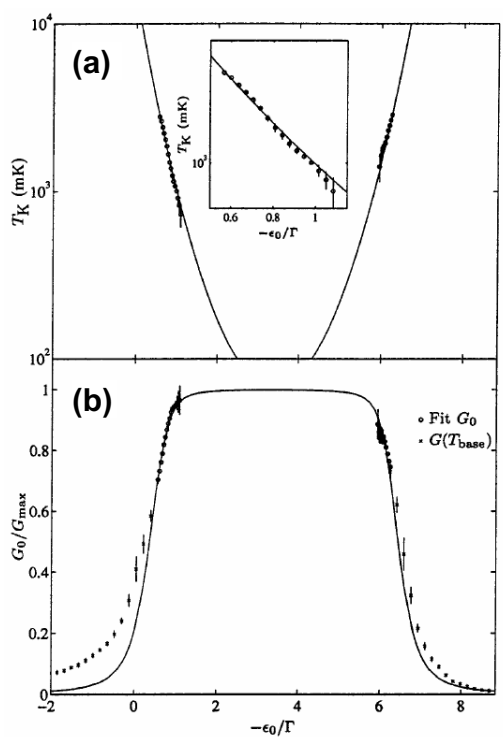

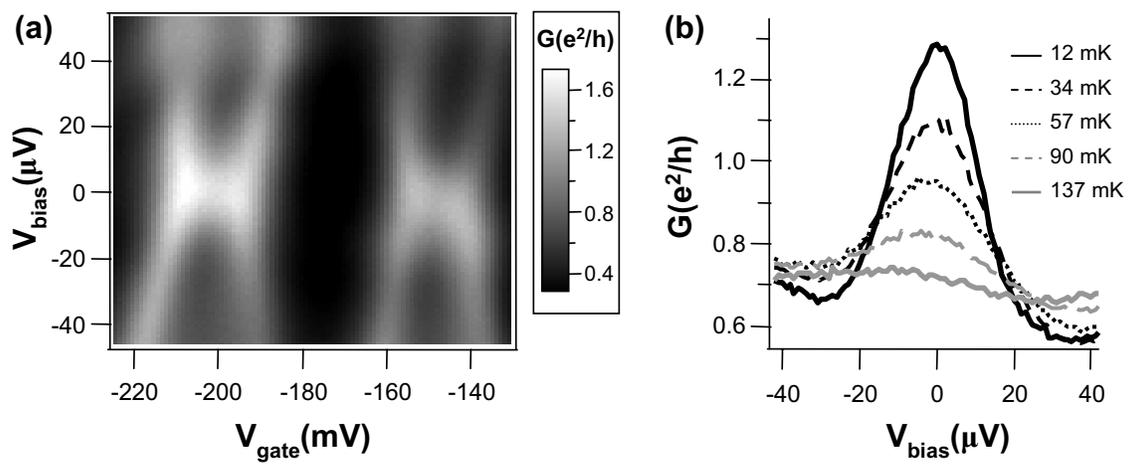

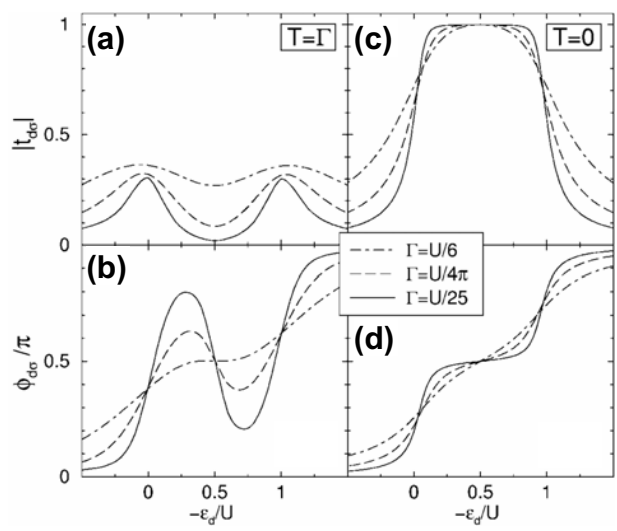

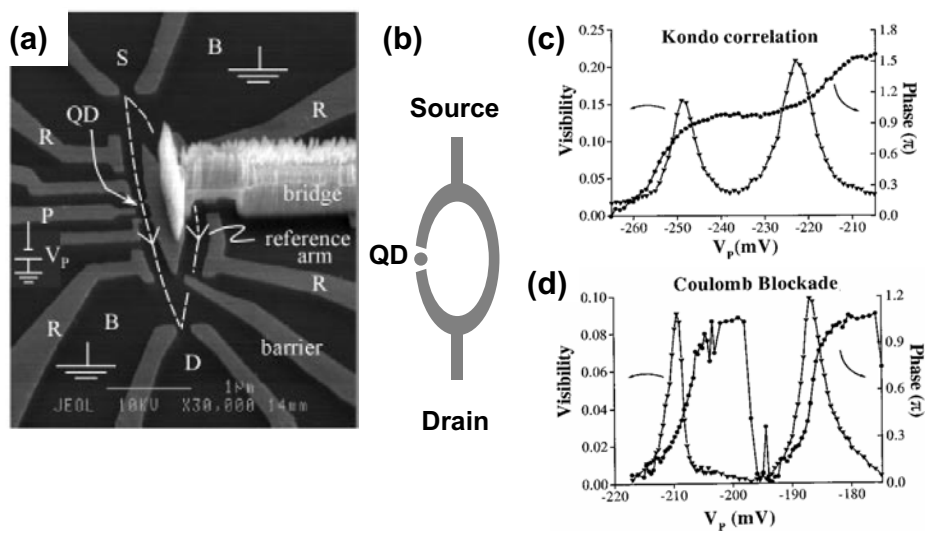

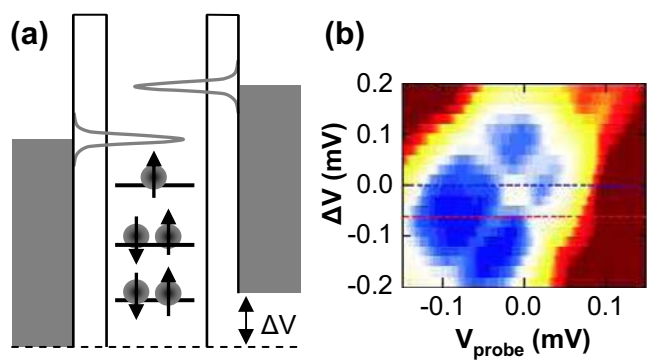

**(a)**

**(b)**

$V_{probe}$ (mV)

$\Delta V$ (mV)

$\Delta V$

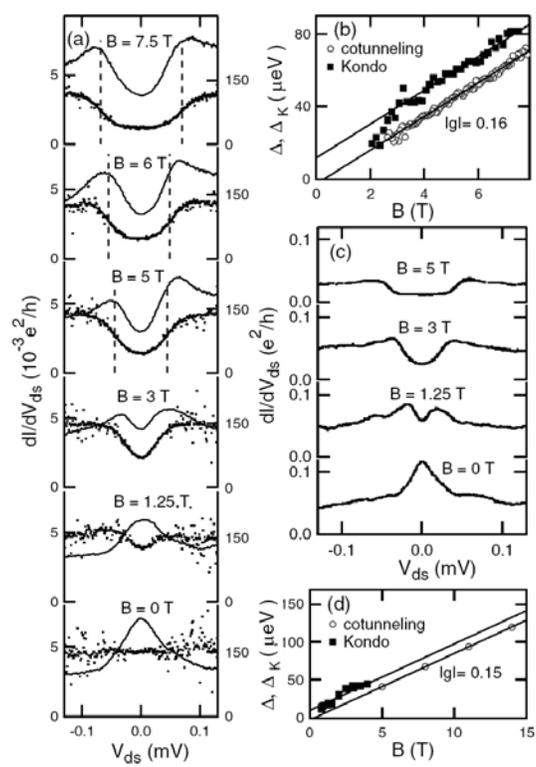

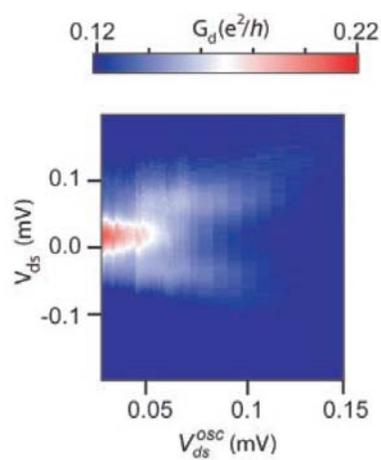

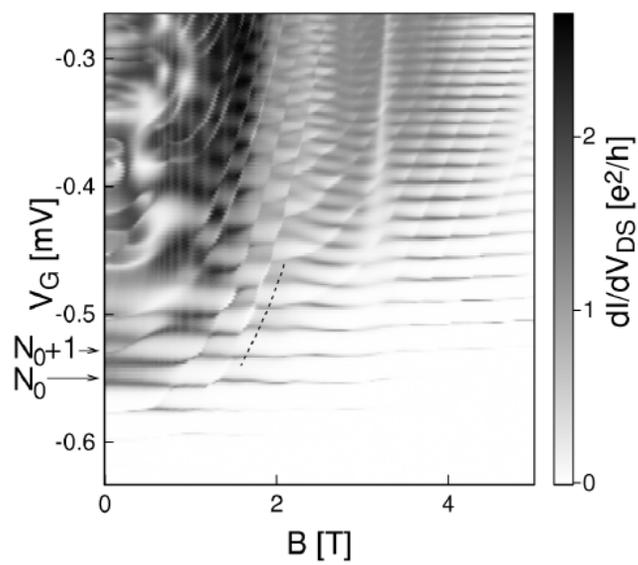

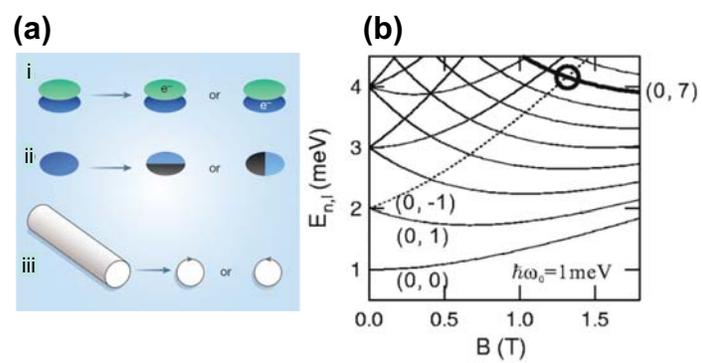

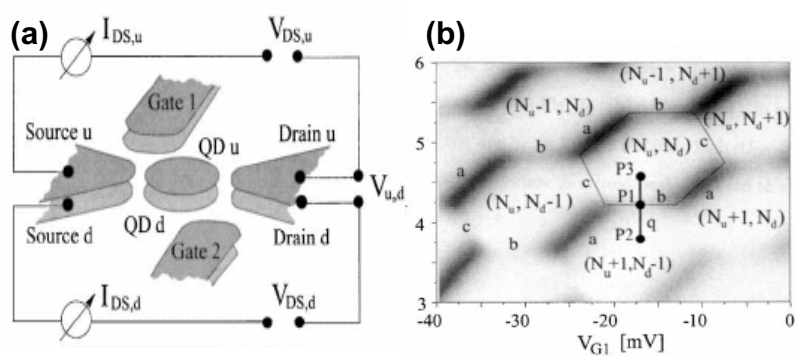

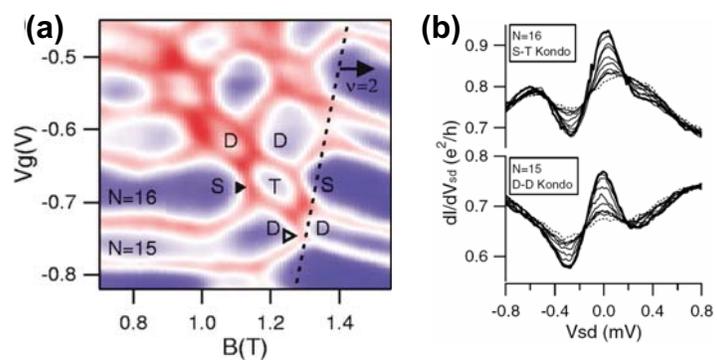

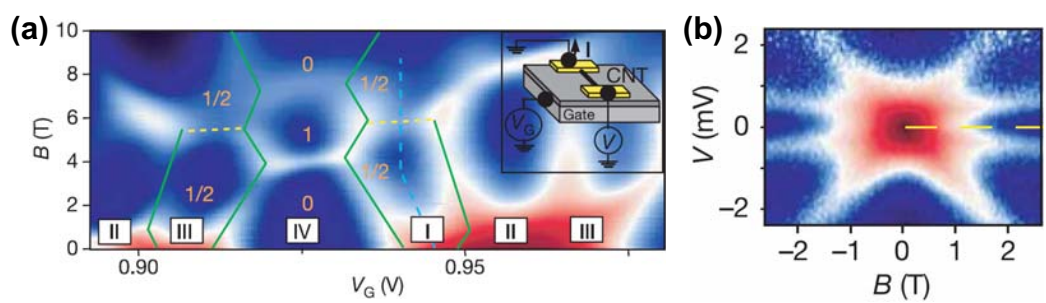

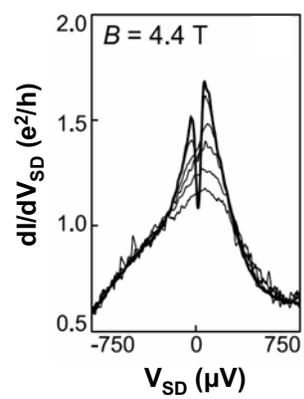